\documentclass[sigconf]{acmart}

\renewcommand\footnotetextcopyrightpermission[1]{}

\usepackage{graphicx}
\usepackage{ulem}
\usepackage{subcaption}
\usepackage{booktabs}
\usepackage{placeins}
\usepackage{float}
\usepackage{amsmath}
\usepackage{tcolorbox}
\usepackage[dvipsnames]{xcolor}

\begin{document} 
\title{Think Harder and Don't Overlook Your Options: Revisiting Issue-Commit Linking with LLM-Assisted Retrieval}
\author{Cole Morgan}
\affiliation{%
  \institution{University of Windsor}
  \city{Windsor}
  \country{Canada}
}
\email{morgan53@uwindsor.ca}

\author{Muhammad Asaduzzaman}
\affiliation{%
  \institution{University of Windsor}
  \city{Windsor}
  \country{Canada}
}
\email{masaduzz@uwindsor.ca}

\author{Shaiful Chowdhury}
\affiliation{%
  \institution{University of Manitoba}
  \city{Winnipeg}
  \country{Canada}
}
\email{shaiful.chowdhury@umanitoba.ca}

\author{Shaowei Wang}
\affiliation{%
  \institution{University of Manitoba}
  \city{Winnipeg}
  \country{Canada}
}
\email{shaowei.wang@umanitoba.ca}

\begin{abstract}
Linking issue reports to the commits that resolve them is essential for software traceability, maintenance, and evolution. Accurate issue-commit links help developers to understand system changes and the rationale behind them. While numerous automated techniques have been proposed, ranging from heuristic and feature-based approaches to modern deep learning and large language model approaches, our goal is to evaluate these techniques to determine which are most effective and efficient. In this study, we revisit several established issue–commit link recovery techniques, including BTLink, EasyLink, FRLink, RCLinker, and Hybrid-Linker, and assess their performance for reranking issue-commit links. We first evaluate different retrieval methods (BM25, BM25L, SBERT-Semantic Search, ANNOY, LSH, HNSW) for their ability to efficiently retrieve relevant commits, reducing the candidate set that must be considered by more computationally expensive models. Using the best retrieval methods, we then investigate the reranking effectiveness of different machine learning-based techniques, including traditional machine learning models, a cross-encoder, and large language models (ChatGPT, Qwen, Gemma, Llama), to refine the reranking of candidate commits and improve precision. Finally, we compare the effectiveness of these techniques. Our results show that dense retrieval methods outperform sparse retrieval approaches in identifying relevant commits and that combining dense and sparse retrieval can improve recall. Additionally, we find that traditional machine learning–based reranking techniques achieve higher performance than LLM-based approaches. Our results highlight that retrieval-based pipelines remain a practical and effective solution for large-scale issue–commit linking, and that simpler models should be carefully considered before adopting computationally expensive LLM-based approaches.
\end{abstract}

\maketitle
\fancyhead{}
\section{Introduction}
Software traceability plays a central role in software maintenance and evolution, allowing developers to follow and understand the relationships between artifacts, such as requirements, design documents, source code, test cases, and issues. Having the capability to trace how these artifacts evolve over time is crucial for understanding how requirements change, design decisions are implemented, and how software systems mature throughout their life cycle~\cite{traceability_trends, Panis}. Such traceability links between artifacts not only support software maintenance and debugging tasks, but also enable developers to comprehend the rationale behind changes, and analyze the evolution and development of system behaviour. As a result, the links between issues and commits are widely analyzed~\cite{traceability} to support a number of activities, such as bug localization, bug assignment, and vulnerability tracking \cite{commitshield}. Links between issues and commits are typically established through manual developer effort, where issue identifiers are explicitly included in commit messages. However, due to the voluntary nature of the task, a large number of commits are missing links to corresponding issues. Thus, it is important to develop automated techniques and tools to establish the missing links between issues and commits~\cite{issuelinkalg}.

Over the years, numerous automated techniques have been proposed to efficiently and accurately link issues with their corresponding bug-fixing commits. These approaches vary widely in their methodologies, from simple keyword-based heuristics \cite{FRLink, HybridLinker} to advanced deep learning models~\cite{Xie, Ruan, BTLink}. Newer approaches employ a two-stage retrieval pipeline: identifying a candidate set of commits that most likely contain the true commit linked to the issue and then reranking those candidates by relevance.

\textbf{Search Space Optimization:} Within this two-stage pipeline, narrowing the candidate commit pool is a critical optimization step. A more refined search space enhances the performance of both retrieval and reranking methods by significantly reducing the noise generated by irrelevant commits. A recent study by Huang et al. \cite{EasyLink} introduced a technique called EasyLink, demonstrating that a one-year temporal window following issue creation allows for a much more comprehensive capture of true links. This stands in contrast to earlier works that often relied on a restrictive seven-day window \cite{Ruan, Bachmann, Nguyen, FRLink, BTLink}. Given an issue, EasyLink uses a one-year time window to generate a candidate set of commits that likely contains the true link. The technique utilizes an information retrieval method to quickly retrieve the potential links for a given issue, and leverages a large language model (LLM) to rerank the commits.

However, we see a gap between what has been done and what we can do. First, it is unclear how realistic the one-year time window is. Does the selection of different datasets lead to different findings? Second, the performance of the retrieval-based issue-commit linking technique depends on the selection of the retrieval method. It is unclear which retrieval method perform best in issue-commit linking. Third, LLM-based retrieval requires specialized hardware for training and inference LLMs, which may not be readily available because of the cost associated with owning the hardware.

This paper presents a study to answer the above mentioned questions. Toward this goal, we consider three different datasets containing issues and commits from several open-source software repositories. We analyze how the selection of the time window affects the ability to uncover true links. To determine how the selection of the retrieval method affects the performance of identifying true links, we consider both dense and sparse retrieval methods. Next, to determine whether we can find alternatives to large language models in reranking retrieved commits, we consider both heuristic and machine-learning models developed in prior studies. Thus, we structure our study considering the following three research questions.

\begin{itemize}
  \item \textbf{How does the selection of time window affect the ability to identify true links?} Our study reveals that while a one-year window from issue creation is effective for specific datasets, it lacks universal applicability. For instance, while this window captures 97\% of true-linked commits in the dataset released alongside the EasyLink technique by Huang et al. \cite{EasyLink}, which we refer to as the EasyLink dataset, its coverage is significantly lower in our other evaluated benchmarks. However, we identify a strong temporal correlation between true links and the issue-closure date. By proposing a hybrid temporal heuristic-combining a one-year window following issue creation with a thirty-day buffer around the issue-closure date, we successfully capture over 97\% of true links across all three datasets.

  \item \textbf{ How does the selection of retrieval method affect the performance of retrieving target commits?} Our results demonstrate that narrowing the search space through temporal filtering significantly improves both precision and recall compared to retrieving candidates from a project's entire commit history. Within these filtered windows, we find that dense vector-based retrieval methods generally outperform traditional sparse methods. Our analysis shows that dense and sparse models retrieve different commits. Using Reciprocal Rank Fusion to combine dense and sparse retrieval methods leverages their complementary strengths, resulting in higher recall than either model alone.

  \item \textbf{ How does the selection of reranking technique affect the performance of the issue-commit link recovery?} The landscape of reranking options for traceability recovery spans a wide spectrum, from traditional machine learning models to modern Transformer-based architectures and Large Language Models. We evaluate these diverse approaches within a reranking environment to determine their relative effectiveness in identifying relevant commits. Our results demonstrate that simpler machine learning models often achieve performance comparable to, or even exceeding, that of state-of-the-art LLMs, emphasizing the importance of evaluating simpler techniques before adopting more resource-intensive solutions.
\end{itemize}

The remainder of this paper is organized as follows. Section~\ref{sec:background} provides the necessary background to contextualize our study. Section~\ref{sec:Datasets} introduces the datasets used in our work, Section~\ref{sec:experimental_setup} details our experimental setup. Section~\ref{sec:research_questions} presents the results and addresses our research questions. Section~\ref{sec:takeways} discusses the implications of our findings. Section~\ref{sec:related_work} reviews related work, and Section~\ref{sec:ThreatsToValidity} outlines the threats to the validity of our study. Finally, Section~\ref{sec:conclusion} concludes the paper.

\section{Background}\label{sec:background}
\subsection{Issue-Commit Linking}
\begin{figure}
    \includegraphics[scale=0.41]{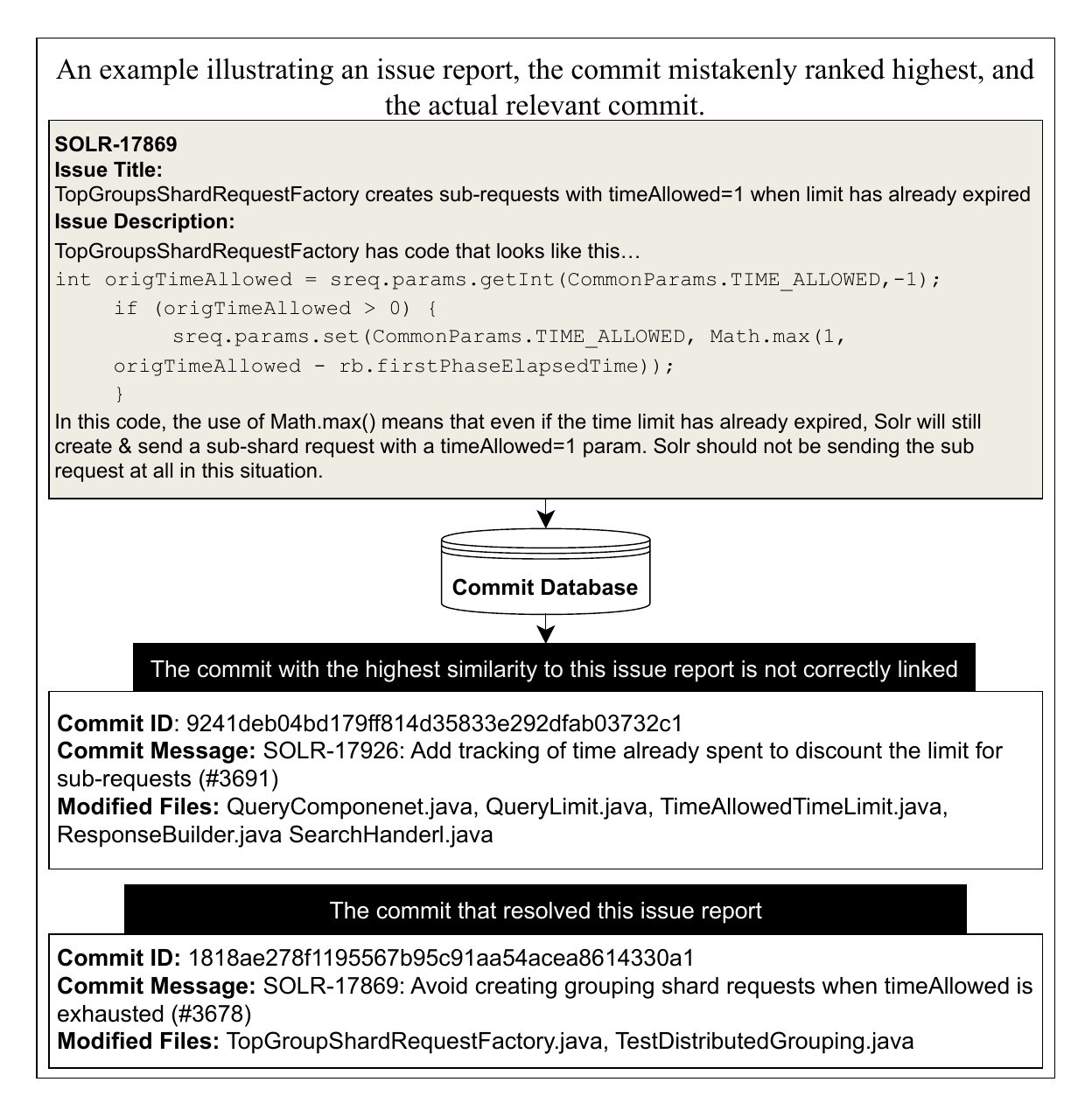}
    \vspace{-0.8cm}
    \caption{An example of a retrieved commit that is similar to but not correlated with the issue report. Linking methods need to be able to identify a relevant commit out of a set of similar items.}
    \Description{A figure showing an example of a retrieved commit that is similar to, but not correlated with, the issue report.}
    \label{issue-commit}
\vspace{-0.5cm}
\end{figure}

Software development projects often rely on issue tracking systems (Jira, GitHub Issues) and version control systems (Git) to manage workflows. Issue tracking systems allow teams to organize tasks, assign priorities, track progress, and allow collaboration among developers, testers, and project managers. Version control systems provide a detailed history of code changes, enabling auditing and coordinated work. Issues can range from bug reports, feature requests, and maintenance tasks, while commits contain the related code changes. Each system captures a different view of the development process. Issues describe what needs to be done and why, often including detailed descriptions, comments, and priority levels. Commits, on the other hand, describe how the changes were implemented, containing code modifications, added or removed files, and a commit message.

Establishing links between issues and commits, known as issue-commit link recovery, allows greater visibility, understanding, and traceability for development artifacts, project maintenance, and analysis. Issue-commit link recovery is the task of automatically establishing these connections when they are missing or incomplete. Ideally, each commit that implements a task or fixes a bug would reference the corresponding issue. However, developers may forget or inconsistently include linked issues in commits. Manually maintaining links between issues and commits is time-consuming. In large software projects, numerous commits and issues may be created daily, making it difficult for developers to consistently reference every relevant issue in their commit messages~\cite{Rodriguez}. This can lead to gaps in traceability, complicating activities such as managing code changes, regression testing, and impact analysis. As shown in Figure~\ref{issue-commit}, the most semantically similar commit may not correspond to the correct bug-fixing commit, highlighting the difficulty of accurately establishing issue-commit links.

Research on issue-commit link recovery is a well-established topic in software engineering. Many automated methods have been proposed, leveraging different sources of information, including textual similarity between issue descriptions and commit messages, metadata features such as authorship, date proximity, and file modifications. However, the task remains challenging. For instance, there can be issues and commits that are semantically similar but unrelated to each other. To address this, more recent studies have implemented deep learning and transformer-based models, and leverage large language models, which are capable of capturing richer semantic and contextual relationships between issues and commits. These approaches aim to improve the accuracy of link recovery by understanding not just keyword overlap but the underlying meaning of both issues and commits.

Additionally, recent work \cite{EasyLink} proposes generating candidate commits using realistic time frames that reflect when true links are likely to occur. Many prior studies rely on balanced datasets or limited candidate selection, which neglect the large number of potential commits and may overestimate technique performance. Models trained on such non-realistic sets of false links may perform differently when evaluated on more realistic candidate sets, underscoring the need to reassess their effectiveness in real-world scenarios.

\FloatBarrier
\begin{figure}[t]
    \includegraphics[scale=0.33]{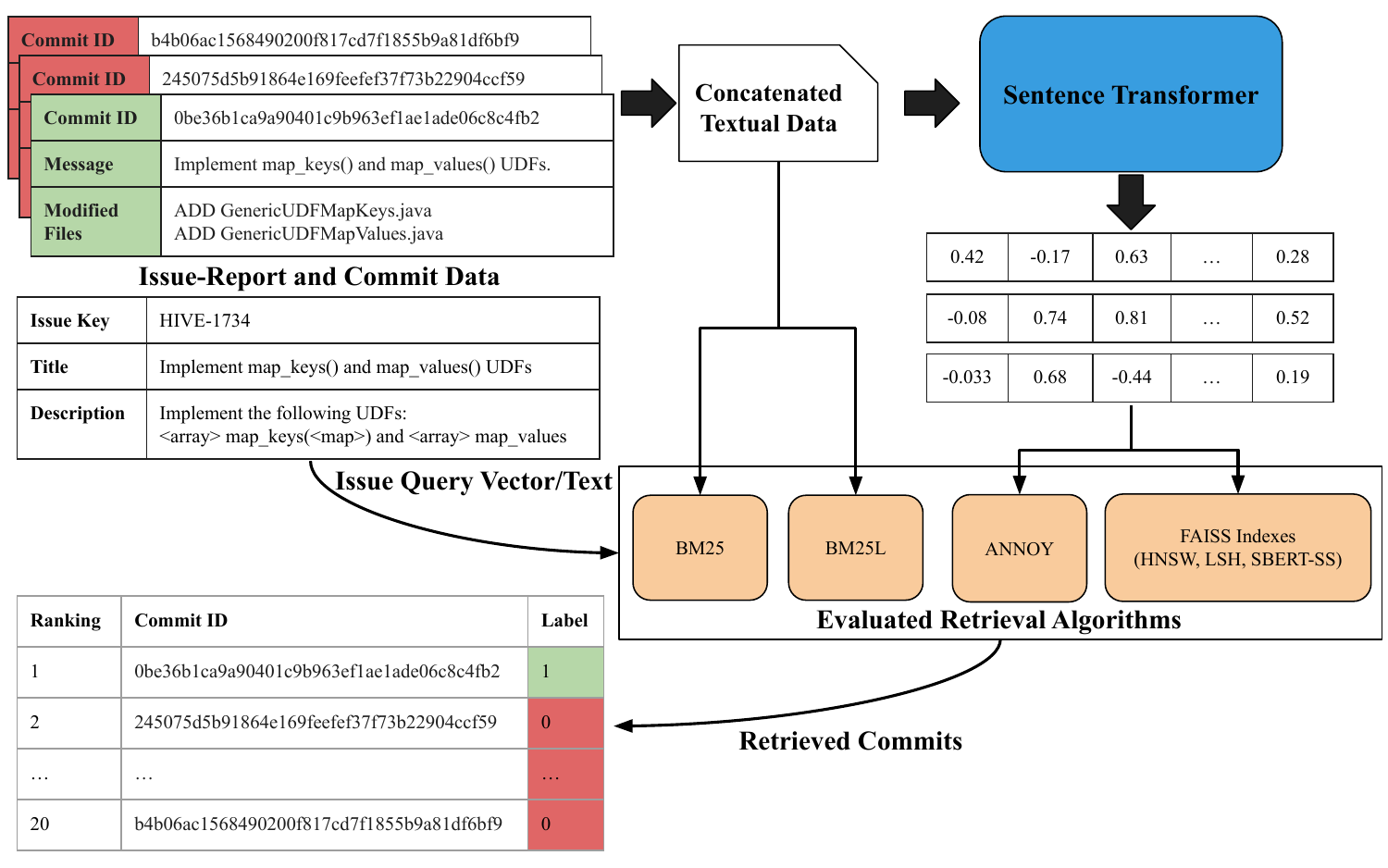}
    \caption{An example of an information retrieval pipeline. Textual data from commits are used to generate sparse/dense vectors allowing for them to be indexed. The data from an issue is than used as a query to retrieve relevant commits.}
    \label{table:retrieval-diagram}
    \Description{An information retrieval pipeline where commit messages are transformed into sparse and dense vector representations and stored in an index. An issue report is then encoded as a query vector, which is used to retrieve the most relevant commits from the indexed repository based on similarity.}
\end{figure}

\subsection{Studied Retrievers}
The choice of retriever can strongly affect both retrieval accuracy and efficiency in issue-commit link recovery, where issues are used as queries to retrieve relevant commits (Figure~\ref{table:retrieval-diagram}). Retrieval methods are commonly divided into sparse and dense retrievers, which differ in how documents are represented and how relevance is computed.

Sparse retrievers represent documents as high-dimensional term vectors and rank results using lexical overlap. We evaluate BM25 \cite{bm25, bm25Improved} and BM25L \cite{bm25l}, both based on TF-IDF style weighting. Term frequency reflects how often a term appears in a document, while inverse document frequency reduces the influence of common terms. Sparse methods are efficient and effective when strong keyword overlap exists, but they depend on exact term matching. In software repositories, issues and commit messages often use different vocabularies, causing vocabulary mismatch.

Dense retrievers instead encode queries and documents into semantic embeddings using transformer-based models. Relevance is then commonly computed using cosine similarity between embeddings, allowing matches beyond exact keyword overlap.

Sparse and dense retrievers differ in whether relevance is determined by lexical overlap or semantic similarity, making their strengths complementary. Reciprocal Rank Fusion (RRF)\cite{RRF} combines their ranked outputs into a unified ranking, weighting documents by rank position rather than raw similarity scores. Documents appearing highly in multiple lists receive higher final ranks.

We evaluated six retrievers previously studied by Pengfei et al. \cite{RAG}: two sparse methods (BM25 and BM25L) and four dense methods. BM25 is widely used for its strong effectiveness and simplicity, while BM25L reduces BM25's sensitivity to document length~\cite{bm25l}.

The dense retrievers evaluated are SBERT Semantic Search (SBERT-SS) \cite{sbert}, Hierarchical Navigable Small World (HNSW) \cite{Hnsw}, Approximate Nearest Neighbours Oh Yeah (ANNOY) \cite{Annoy,ANN}, and Locality-Sensitive Hashing (LSH) \cite{lsh}. SBERT-SS performs exhaustive cosine similarity search over all document embeddings, maximizing retrieval accuracy at the cost of high computational complexity. In contrast, HNSW, ANNOY, and LSH are approximate nearest neighbor (ANN) methods that improve efficiency by avoiding exhaustive comparisons. These methods use specialized index structures to quickly identify a small set of candidate vectors that are likely to be similar to the query, and compute similarity only within this subset. While this approximation may lead to a slight reduction in retrieval effectiveness, it significantly reduces query time, making ANN methods suitable for large-scale retrieval settings. HNSW organizes embeddings in a multi-layer graph, ANNOY uses random projection trees, and LSH maps similar vectors into shared hash buckets.

\section{Dataset Description}\label{sec:Datasets}
For the purpose of our evaluation, we considered the datasets provided by Lan et al. \cite{BTLink} (i.e., the BTLink dataset, named after the BTLink technique proposed in that work) and Huang et al. \cite{EasyLink} (i.e., the EasyLink dataset, named after the EasyLink technique proposed in that work). The BTLink dataset consists of 13 Apache projects and one popular GitHub project, covering different popular programming languages. This dataset contains 39,051 true links, 18,787 issues, and 36,199 commits collected from 2004--2022. The EasyLink dataset consists of 20 open-source projects, 159,592 true links, 54,922 issues, and 346,501 commits collected from 2003--2020. These projects use a mixture of GitHub and Jira for their issue reporting service. Both datasets contain issue reports associated with each project, commits that changed files, and a list of changed/modified/deleted files and code diff. Since we evaluate a range of methods requiring different types of information and features, we recollected all issues and commits from both datasets to ensure that all necessary data was available. Apache Airflow, contained in the EasyLink dataset, has changed from Jira to GitHub in recent years, making all of its older Jira issue reports unavailable. For this reason, we excluded the project from our analysis.


To increase the generalizability of our findings, we prepared an additional dataset. We collected ten popular Apache projects that use Jira as the issue management system, are not included in the previous datasets, and contain the largest number of issue reports. Our selected Apache projects are Spark, Ozone, HBase, Hive, Camel, Kafka, Solr, Hadoop Common, NiFi, and Impala. This dataset referred to as the Apache dataset, consists of 208,858 true links, 130,538 issue reports, and 166,460 commits collected from 2006--2025.  


For each selected project, we collected all available issue reports using the Jira API. We used the PyDriller library to collect all the commits~\cite{PyDriller_2018}. For each commit, we collect the hash, author, date, message, names of all modified files, diff of all modified files, and names of modified methods in each file.

For identifying issue-fixing commits, we applied several filtering steps. First, we removed commits that had more than one parent to remove duplicate commits created when merging branches and pull requests. Next, we excluded commits that did not contain any file changes, as such commits are unlikely to contain substantial modifications or bug fixes. Finally, to increase the likelihood that retained commits are bug-fixing commits, we considered only those which contain an issue key (e.g., PROJECT-123) in their commit message for Jira projects. This gave us a collection of commits that are strong candidates for being issue-fixing commits, as they contain code changes and explicitly reference corresponding issues. These references between commits and issues are used to construct the set of true links, assuming that developers who explicitly mention an issue are intentionally linking the commit to it. To verify that our true links collected using this method are reliable and are correctly linked, we randomly selected 40 issue-commit links from each project, totalling 400 links. The random selection was made to ensure that we achieved a minimum of 95\% confidence level and 5\% confidence interval. The first two authors of the study manually checked each link to see if the commit was correctly linked to its corresponding issue report. In the case of disagreements they discussed with each other to resolve the disagreement. We computed Cohen’s kappa~\cite{Cohen1960ACO} to measure inter-rater agreement before the disagreement was resolved. We obtained a kappa value of 0.93 which indicates a high degree of agreement. Through this study, we found that 13 of 400 issues sampled from our Apache Dataset were incorrectly linked to their issue-fixing commits. This occurred when commits referenced similar issues when fixing or implementing a different one, as well as commits that merged branches or multiple commits together referencing the issues that were fixed in the previous commits. Since our constructed datasets match closely with the manual labels (i.e., ground-truth), our dataset is of high quality.

\section{Experimental Setup}\label{sec:experimental_setup}

\subsection{Experiment Configuration}
Before evaluating retrieval and ranking performance, we first analyzed time frames between real-world true links to identify temporal boundaries that capture the most true links. Previous work has varied from restrictive 7-day windows to expansive one-year periods. Our experiments tested combinations relative to both issue creation and closure dates to determine a window that maximizes true link inclusion across the dataset.

Following the establishment of an optimal temporal window, we evaluated retrieval methods for identifying relevant commits. Retrieval reduces the volume of data processed downstream, especially in large repositories, by ranking commits to create a manageable candidate set. We compared multiple algorithms across varying $K$-values, where $K$ denotes the number of top-ranked candidate commits returned by the retriever for subsequent reranking. We evaluated $K \in \{1, 5, 10, 20, 50\}$ to assess the trade-off between recall and candidate set size. The best-performing method, based on metrics like Recall@K, served as the candidate generation stage for subsequent experiments. High recall is critical, as insufficient retrieval cannot be compensated for by later reranking, which primarily improves precision for top-ranked candidates. To mitigate potential data leakage, we also removed all issue identifiers (issue keys) from commit messages, ensuring that no explicit links between commits and issues were available to either the retrieval or reranking methods.

Using this selected retriever, we evaluated the effectiveness of several previously proposed machine learning–based techniques designed for classifying issue-commit links. Specifically, we examined BTLink~\cite{BTLink}, FRLink \cite{FRLink}, RCLinker \cite{RCLinker}, and Hybrid-Linker~\cite{HybridLinker}. These models differ in their feature representations, architectures and learning strategies, ranging from combining pre-trained transformer models leveraging semantic information to random forests and gradient boosting on combinations of textual and non-textual data. Additionally, cross-encoder–based reranking \footnote{https://huggingface.co/cross-encoder/ms-marco-MiniLM-L6-v2} was included as a strong baseline due to its proven effectiveness in information retrieval and retrieval-augmented generation.

Finally, we compared these reranking techniques with Large Language Models, which have shown strong performance on software-related text and may identify issue–commit links without task-specific training. We evaluated GPT-5.1 via the OpenAI API and locally run Llama-3.1-8B, Gemma-7B, and Qwen3-32B to compare their effectiveness.

\textbf{Train/Test Split:} For all reranking experiments, we partitioned each project in a dataset chronologically, assigning the newest 20\% of issue reports as the test set and using the remainder for training and validation. This temporal split ensures that our evaluation reflects a realistic scenario where models are trained on historical data and evaluated on future issues. For projects where the test set contains more than 1,000 issue reports, we randomly sampled 1,000 issues and repeated this process five times, reporting the average results to mitigate sampling bias.

\subsection{Evaluation Metrics}
We adopt the metrics used by Huang et al. \cite{EasyLink}, defined as follows.

\textbf{Precision@K (P@K)} measures the proportion of relevant items among the top-$K$ retrieved results:
\begin{equation}
\text{Precision@}K=\frac{1}{N}\sum_{i=1}^{N}\frac{Rel_i}{K}
\end{equation}
where $N$ is the number of test cases and $Rel_i$ is the number of correctly linked commits among the $K$ retrieved results.

\textbf{Hit@K} measures the fraction of queries where at least relevant commit appears in the top-$K$ results:
\begin{equation}
\text{Hit@}K = \frac{1}{N} \sum_{i=1}^{N} \mathbb{I}\left(\text{rank}_i \leq K\right)
\end{equation}

\textbf{Recall@K} measures the fraction of all relevant commits successfully retrieved within the top-$K$:
\begin{equation}
\text{Recall@}K=\frac{1}{N}\sum_{i=1}^{N}\frac{Rel_i}{TotalRel_i}
\end{equation}

\textbf{NDCG@K} evaluates ranking quality by weighting relevant items higher when they appear earlier, normalized so that a perfect ranking scores 1.0:
{\small
\begin{equation}
DCG_p=\sum_{i=1}^{p}\frac{2^{rel_i}-1}{\log_2(i+1)}, \quad
NDCG_p=\frac{DCG_p}{IDCG_p}, \quad
IDCG_p=\sum_{i=1}^{|REL_p|}\frac{2^{rel_i}-1}{\log_2(i+1)}
\end{equation}
}
\textbf{Mean Reciprocal Rank (MRR)} measures how early the first relevant result appears, averaged across all queries:
\begin{equation}
MRR=\frac{1}{N}\sum_{i=1}^{N}\frac{1}{rank_i}
\end{equation}

\section{Answers to Research Questions}\label{sec:research_questions}

\subsection{RQ1: Selection of the Time Window}

\begin{figure*}[!t]
    \begin{subfigure}{0.33\textwidth}
        \includegraphics[width=\linewidth, keepaspectratio]{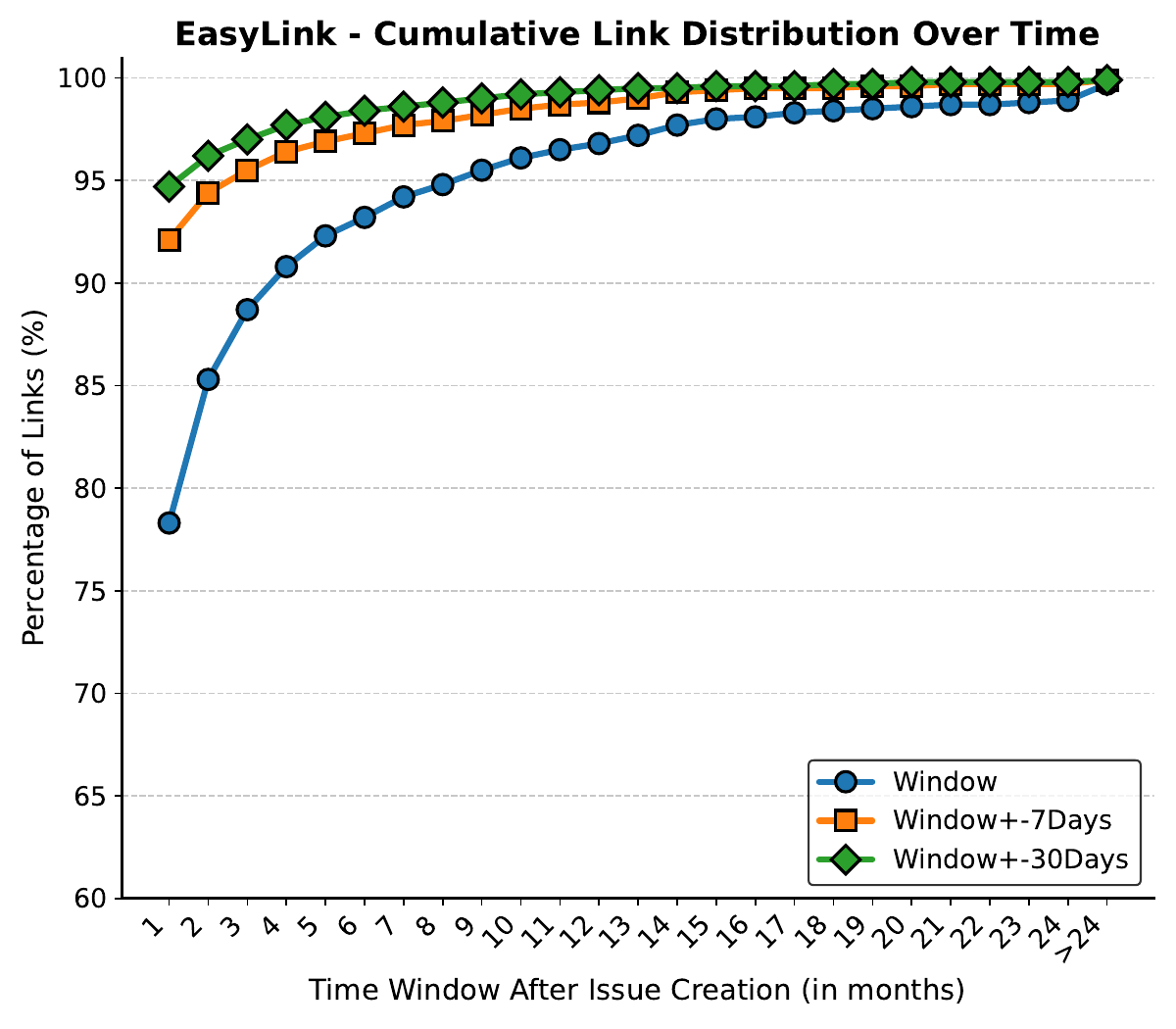}
    \end{subfigure}
    \begin{subfigure}{0.33\textwidth}
        \includegraphics[width=\linewidth, keepaspectratio]{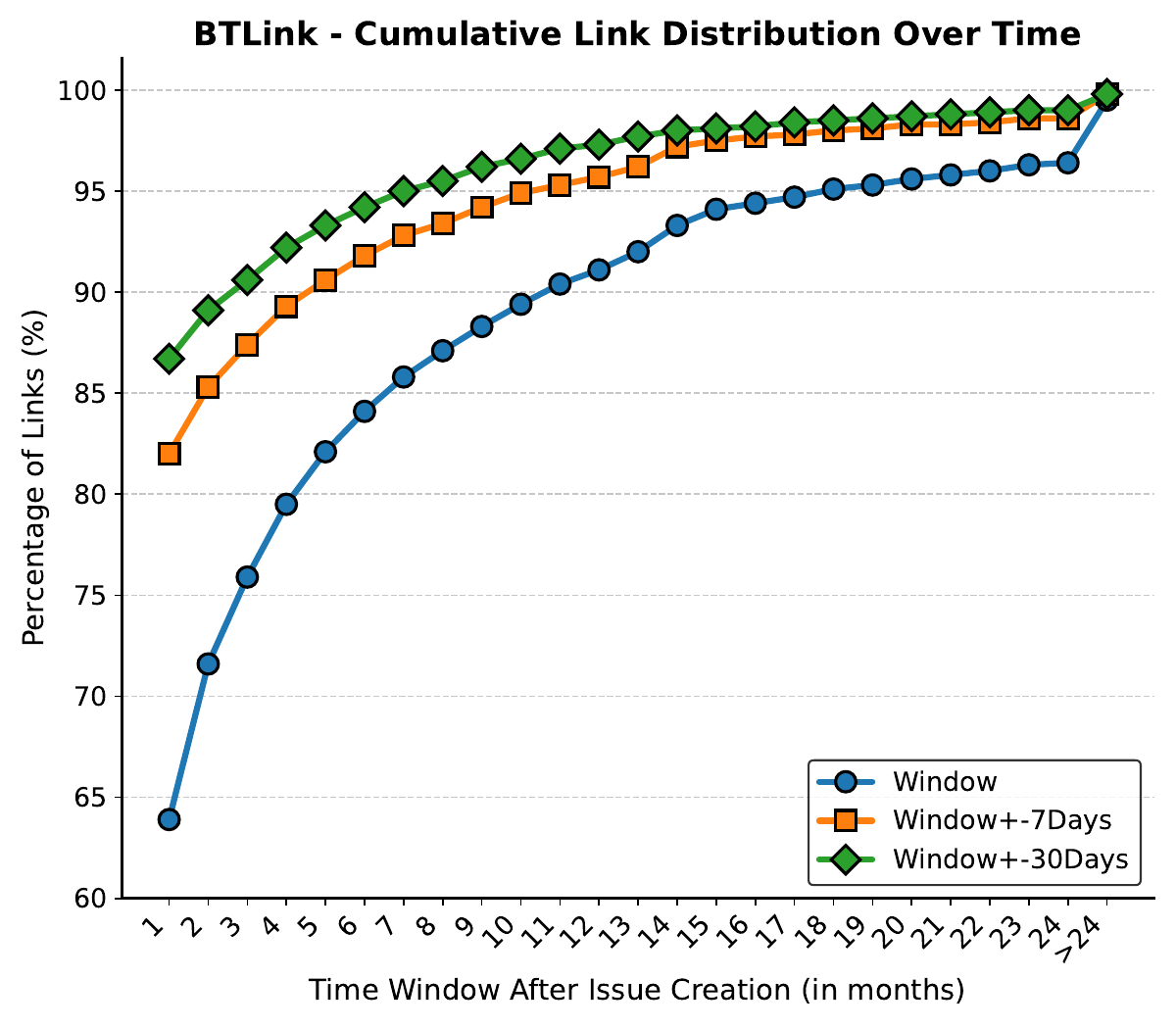}
    \end{subfigure}
    \begin{subfigure}{0.33\textwidth}
    \includegraphics[width=\linewidth, keepaspectratio]{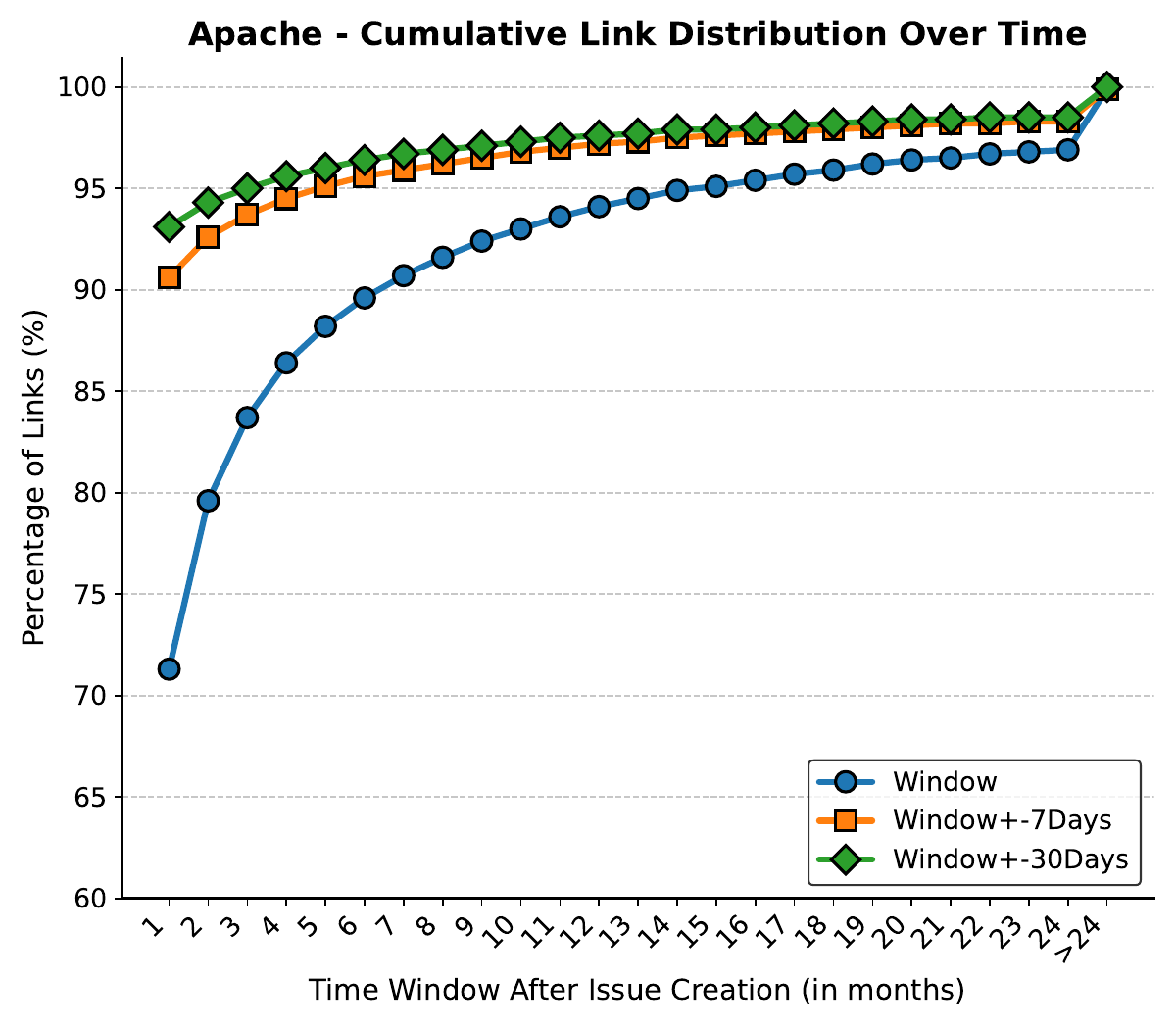}
        \end{subfigure}
    \caption{Distribution Of True Links Captured by Time Frames on Our Datasets}
    \label{rq1_fig}
    \Description{A set of three plots (EasyLink, BTLink, and Apache) showing the percentage of ground-truth issue-commit links successfully captured under three different temporal window configurations. The figure compares how varying time window constraints impact true link coverage across datasets.}

\end{figure*}


\textbf{Motivation:} To recover missing link(s) for a given issue, we define a temporal window that determines a candidate commit pool for each issue. Commits outside this window are excluded from all downstream steps, making the window a hard upper bound to search for missing link(s). The larger the number of candidate commits for an issue, the difficult it is to find the target commit(s) due to a large search space. A small temporal window can increase the risk of missing the target commit(s) due to the presence of a small number of commits. This RQ studies the trade-off between window size and true link coverage across datasets, with the goal of identifying a window that balances high coverage and practical feasibility.


\textbf{Approach}: To answer this research question, we considered three different windows. These are a) one year window after issue creation, b) one year window and seven days before and after issue closure, and c) one year window and thirty days before and after issue closure. Huang et al. found that 97\% of fix commits occurred within the one year time window for the EasyLink dataset~\cite{EasyLink}. Lan et al. considered seven days before or after the issue creation date and the closed date to identify potential links likely to contain the fix commits for an issue. We did not consider the seven-day time window because a prior study found that only 59\% \cite{EasyLink} of issues had a corresponding fix commit within seven days of their creation. However, we did consider seven days and thirty days before and after issue closure and combined them with the one-year time window to determine how combining two different windows affects the ability to retrieve relevant commits.

\textbf{Results}: Figure~\ref{rq1_fig} shows the percentage of true links that can be found considering different time windows for the three datasets. Our results show that the one-year time window collects approximately 97\% of all true links for the EasyLink dataset; it only collects 91.6\% and 94.1\% of true links for the BTLink and Apache datasets, respectively. In contrast, when applying the traditional seven-day time window before and after issue closure, we found that 87.6\% of true links were collected for the EasyLink dataset and 74.7\% of true links were collected for the BTLink dataset. The number increases to 87.5\% for the Apache dataset. Our evaluation finds that neither time frame is optimal across all datasets. While the one-year window maximizes recall by capturing nearly all true links, it also increases the number of candidate commits that must be considered and may reduce precision. Contrasting this, the seven-day window leverages both the issue-creation and issue-closure date, allowing links to be identified within a short period surrounding either event. While issue closure typically indicates that an issue has been resolved, in practice, developers may complete multiple tasks before closing an issue, or commits may be made on separate branches and merged later. In our datasets, 14.3\% of true links in EasyLink dataset occur after issue closure (11.7\% in seven days), 12.2\% in BTLink dataset (7.7\% within seven days), and 13.2\% (7.4\% within seven days) for our Apache dataset. In contrast, commits before issue creation are rare: only 0.3\% of links in EasyLink dataset, 0.5\% of links in BTLink dataset, and 0.1\% of links in Apache dataset. Using both issue creation and closure dates, we analyzed how link coverage varies across time frames. We combined a variable post-creation window with a fixed seven-day pre and post-closure window to evaluate when 95\%, 99\%, and 100\% of links are captured.
\begin{itemize}
    \item \textbf{EasyLink Dataset:} 95.3\% of true links appear within 3 months plus seven days around issue closure; 99\% within 13 months, 99.5\% within 16 months, and 99.9\% \sout{occur after} within 38 months.
    \item \textbf{BTLink Dataset:} 95.3\% of true links appear within 11 months; 99\% within 27 months, 99.5\% within 39 months, and 99.8\% within 74 months.
    \item \textbf{Apache Dataset:} 95.1\% of true links appear within 5 months; 99\% within 35 months, 99.5\% within 54 months, and 99.9\% within 88 months.
\end{itemize} 
These results indicate that achieving near-complete coverage (99–100\%) requires significantly long time frames. In practice, a balance between coverage and feasibility is necessary, as attempting to capture every link can introduce a large number of false links that can negatively impact the performance of issue-commit linking techniques. For our subsequent research questions and evaluations, we adopted a time window consisting of one year after issue creation, combined with a thirty-day window before and after issue closure. Such a time window captures 97\% of true links across all of our datasets.




\subsection{RQ2: Selection of the Retrieval Method}


\textbf{Motivation:} This research question focuses on understanding how the choice of retrieval algorithm affects the ability to find relevant commits given an issue report. In particular, we examined the trade-offs between sparse and dense retrieval approaches, the impact of approximate nearest-neighbour indexing structures, and the extent to which semantic representations improve retrieval effectiveness over lexical matching. This can help us to select the correct retrieval algorithm. The goal of the retrieval method is to quickly identify $top-K$ candidate commits for a given issue that most likely contain the target commit. A more accurate but time consuming reranking technique is then applied to rank the relevant commit(s) in the top position(s). If a relevant commit is not retrieved within the $top-K$ candidate commits, it cannot be recovered by the reranking technique. Therefore, high recall at low $K$ values is important. By using retrieval algorithms to narrow down the candidate commits for a given issue, we reduce the search space from thousands of potential commits to a small set of the most promising candidates, enabling downstream reranking techniques to operate more efficiently and focus on the most relevant commits retrieved.

\begin{table*}[t]
\centering
\small
\setlength{\tabcolsep}{4pt}
\begin{tabular}{lccc|ccccc|ccccc|cc}
\toprule
Retriever & Dataset & P@1 & R@1 & P@10 & Hit@10 & R@10 & MRR@10 & NDCG@10 &
P@20 & Hit@20 & R@20 & MRR@20 & NDCG@20 & R@30 & R@50 \\
\midrule
ANNOY & EasyLink & 0.623 & 0.530 & 0.109 & 0.850 & 0.808 & 0.708 & 0.715 & 0.058 & 0.883 & 0.843 & 0.710 & 0.725 & 0.864 & 0.882 \\
BM25 & EasyLink & 0.505 & 0.431 & 0.091 & 0.735 & 0.698 & 0.587 & 0.601 & 0.050 & 0.788 & 0.749 & 0.591 & 0.614 & 0.772 & 0.800 \\
BM25L & EasyLink & 0.191 & 0.165 & 0.061 & 0.512 & 0.485 & 0.287 & 0.331 & 0.038 & 0.615 & 0.584 & 0.294 & 0.357 & 0.640 & 0.707 \\
HNSW & EasyLink & 0.609 & 0.517 & 0.107 & 0.827 & 0.788 & 0.690 & 0.699 & 0.057 & 0.850 & 0.815 & 0.692 & 0.706 & 0.832 & 0.846 \\
LSH & EasyLink & 0.169 & 0.143 & 0.052 & 0.439 & 0.399 & 0.248 & 0.274 & 0.032 & 0.526 & 0.481 & 0.254 & 0.296 & 0.536 & 0.610 \\
RRF & EasyLink & 0.620 & 0.529 & 0.109 & 0.861 & 0.819 & 0.707 & 0.719 & 0.059 & \textbf{0.903} & \textbf{0.865} & 0.710 & 0.732 & \textbf{0.886} & \textbf{0.907} \\
SBERT-SS & EasyLink & \textbf{0.650} & \textbf{0.556} & \textbf{0.111} & \textbf{0.863} & \textbf{0.824} & \textbf{0.726} & \textbf{0.734} & \textbf{0.060} & 0.897 & 0.861 & \textbf{0.729} & \textbf{0.744} & 0.883 & 0.902 \\
ANNOY & BTLink & \textbf{0.780} & \textbf{0.671} & \textbf{0.119} & \textbf{0.932} & \textbf{0.883} & \textbf{0.835} & \textbf{0.824} & \textbf{0.063} & \textbf{0.952} & \textbf{0.909} & \textbf{0.836} & 0.831 & \textbf{0.925} & 0.940 \\
BM25 & BTLink & 0.658 & 0.565 & 0.100 & 0.813 & 0.756 & 0.706 & 0.694 & 0.054 & 0.863 & 0.809 & 0.709 & 0.708 & 0.833 & 0.860 \\
BM25L & BTLink & 0.392 & 0.338 & 0.085 & 0.710 & 0.658 & 0.491 & 0.517 & 0.049 & 0.783 & 0.732 & 0.496 & 0.537 & 0.774 & 0.819 \\
HNSW & BTLink & 0.770 & 0.663 & 0.117 & 0.916 & 0.868 & 0.824 & 0.812 & 0.062 & 0.933 & 0.891 & 0.825 & 0.819 & 0.903 & 0.914 \\
LSH & BTLink & 0.339 & 0.289 & 0.081 & 0.682 & 0.630 & 0.441 & 0.471 & 0.048 & 0.768 & 0.717 & 0.447 & 0.494 & 0.766 & 0.823 \\
RRF & BTLink & 0.758 & 0.650 & 0.114 & 0.907 & 0.852 & 0.807 & 0.792 & 0.062 & 0.941 & 0.894 & 0.809 & 0.804 & 0.917 & 0.937 \\
SBERT-SS & BTLink & \textbf{0.780} & \textbf{0.671} & \textbf{0.119} & \textbf{0.932} & \textbf{0.883} & \textbf{0.835} & \textbf{0.824} & \textbf{0.063} & \textbf{0.952} & \textbf{0.909} & \textbf{0.836} & \textbf{0.832} & \textbf{0.925} & \textbf{0.941} \\
ANNOY & Apache & 0.750 & 0.682 & 0.102 & 0.892 & 0.849 & 0.800 & 0.792 & 0.053 & 0.910 & 0.871 & 0.802 & 0.798 & 0.882 & 0.893 \\
BM25 & Apache & 0.690 & 0.630 & 0.091 & 0.817 & 0.773 & 0.731 & 0.721 & 0.048 & 0.851 & 0.808 & 0.733 & 0.730 & 0.828 & 0.847 \\
BM25L & Apache & 0.241 & 0.220 & 0.060 & 0.552 & 0.519 & 0.336 & 0.371 & 0.036 & 0.647 & 0.610 & 0.342 & 0.395 & 0.664 & 0.724 \\
HNSW & Apache & 0.740 & 0.673 & 0.100 & 0.875 & 0.835 & 0.789 & 0.780 & 0.052 & 0.891 & 0.854 & 0.790 & 0.786 & 0.863 & 0.871 \\
LSH & Apache & 0.221 & 0.200 & 0.053 & 0.491 & 0.455 & 0.299 & 0.326 & 0.032 & 0.580 & 0.540 & 0.305 & 0.348 & 0.594 & 0.658 \\
RRF & Apache & \textbf{0.780} & \textbf{0.710} & \textbf{0.105} & \textbf{0.924} & \textbf{0.880} & \textbf{0.827} & \textbf{0.818} & \textbf{0.055} & \textbf{0.952} & \textbf{0.913} & \textbf{0.829} & \textbf{0.827} & \textbf{0.923} & \textbf{0.934} \\
SBERT-SS & Apache & 0.768 & 0.699 & \textbf{0.105} & 0.920 & 0.878 & 0.822 & 0.815 & \textbf{0.055} & 0.940 & 0.902 & 0.823 & 0.822 & 0.915 & 0.928 \\
\bottomrule
\end{tabular}
\caption{Performance of Different Retrieval Methods on our Datasets}
\label{table:rq2_results}
\vspace{-0.6cm}
\end{table*}

\textbf{Approach:} This section discusses how we conducted the evaluation to answer the research question. Given an issue, we indexed all commits that appeared within the selected time window using an information retrieval system. Next, we used the issue as a query to the information retrieval system to quickly locate a small set of commits that contained the fix commit(s). Each commit is represented by a textual description combining the commit message and structured code terms. Code terms list each modified file with its modification type (added, removed, or modified) and the names of all changed methods. Queries are formed by concatenating the issue title and description, with issue keys removed to prevent lexical matches and data leakage. For each issue, we considered the one-year window and thirty days before and after the issue closure to obtain plausible commits. This temporal filter reduces unlikely matches and focuses on commits more likely related to the issue. The next step was to store these commits in an information retrieval system. For dense retrieval methods, we generated dense vector embeddings using the sentence transformer all-MiniLM-L12-v2\footnote{https://huggingface.co/sentence-transformers/all-MiniLM-L12-v2}. This model was selected due to its strong performance and popularity. Although its smaller counterpart, all-MiniLM-L6-v2\footnote{https://huggingface.co/sentence-transformers/all-MiniLM-L6-v2}, is the most widely downloaded model on Hugging Face, the larger L12 variant has been shown to achieve superior performance in similarity detection. 

The metrics for our evaluation are Recall@K, Precision@K, Hit@K, MRR@K, and NDCG@K. We retrieve up to $K=50$ results for each project and calculate metrics for $K=[1, 5, 10, 20, 30, 50]$. This range of values allows us to find the optimal $K$-Value for this task. A larger amount of retrieved documents can yield a higher recall however this also introduces more negative data which can make it harder to correctly identify relevant commits. Sparse retrieval was implemented using the rank-bm25 \cite{bm25} library with the BM25Okapi and BM25L methods. Dense retrieval indexes are built using the FAISS \footnote{https://github.com/facebookresearch/faiss} and ANNOY \footnote{https://github.com/spotify/annoy} Libraries. SBERT-SS, HNSW, and LSH are implemented in FAISS using the IndexFlatIP, IndexHNSWFlat, and IndexLSH indexes, respectively.

\textbf{Results:} From our evaluation, we find that dense vector-based methods perform the best at retrieving relevant commits to an issue (Table~\ref{table:rq2_results}). These results suggest that capturing semantic similarity, rather than relying solely on lexical overlap, is important for accurately retrieving relevant commits. Our results find that SBERT-SS \cite{sbert} achieves the highest precision and recall for all $K$-Values, with R@10 being 82.4\%, 88.3\%, and 87.8\%. Our approximate nearest neighbour algorithms, ANNOY and HNSW, have comparable results, only being a couple of percent lower. ANNOY \cite{Annoy} in particular has shown that it can match SBERT-SS results on the BTLink dataset. This result suggests that ANN algorithms are able to successfully detect and retrieve relevant commits for this task. However, their speed optimizations cause some relevant commits to be missed. ANN algorithms are more efficient and can be better options in time-constrained tasks where results are required to be retrieved as fast as possible. Additionally, we find that BM25L and LSH perform very poorly compared to other algorithms tested (Table~\ref{table:rq2_results}). LSH, despite also using dense embeddings, performs noticeably worse. This outcome suggests that simply using dense representations is insufficient; the indexing structure may be equally important. LSH \cite{lsh} groups vectors into hash-based buckets, which can limit its ability to capture fine-grained similarity and significantly reduce retrieval accuracy compared to other methods. The poor performance of BM25L suggests that longer documents are more likely to share strong keyword overlap with unrelated documents. By altering how document length is handled compared to BM25, BM25L \cite{bm25l} increases the retrieval of irrelevant results.

\textbf{Reciprocal Rank Fusion Retrieval:} Although SBERT-SS and BM25 perform well individually, their different representations lead them to retrieve different true links. We combined their results using Reciprocal Rank Fusion (RRF) to leverage their complementary strengths. Each retrieved commit is assigned a rank-based score computed as ($\text{RRF Score} = \frac{1}{60 + \text{rank}}$). If a commit is retrieved by both methods, their scores are summed. The top 50 commits ranked by RRF score form the final candidate set. We found that this hybrid sparse-dense approach does not retrieve as good results at $K=10$. However, this technique yields small but consistent improvements in recall at $K$ = 20 as shown in Table~\ref{table:rq2_results}.

In summary, the selection of retrieval method plays a critical role in the effectiveness of information retrieval systems, particularly for issue–commit link recovery. Our findings demonstrate that dense embedding–based retrieval methods consistently outperform sparse approaches, increasing recall@10 by more than 10\%. Approximate nearest neighbour (ANN) methods, such as HNSW and ANNOY, offer a trade-off between effectiveness and efficiency. These methods are able to achieve retrieval performance comparable to SBERT-SS, which compares the query with all document vectors, while offering improved retrieval efficiency, making them well-suited for large-scale or time-constrained systems. Importantly, our results also show that sparse retrieval methods should not be disregarded, while sparse approaches such as BM25 are limited by vocabulary mismatch. By combining sparse and dense retrieval methods using Reciprocal Rank Fusion, we are able to leverage the strengths of both representations, yielding improved recall at higher $K$ values.


\subsection{RQ3: Selection of The Reranking Technique}

\textbf{Motivation}: While retrieval methods focus on identifying a candidate set of potentially relevant commits for a given issue, reranking techniques determine the final ordering of these candidates and therefore have a direct impact on issue-commit link recovery. The goal of this research question is to determine how the selection of reranking techniques affects performance in recovering issue-commit links compared with traditional machine learning, deep learning, and large language model-based techniques. In particular, we examined whether increased model complexity translates into measurable improvements in reranking quality.

\textbf{Approach:} Based on RQ2, we used the BM25 + SBERT-SS Reciprocal Rank Fusion~\cite{RRF} (RRF) retrieval method, which achieves the highest recall, to generate candidate commit sets for all reranking experiments. We split the data chronologically, assigning the newest 20\% of issues as the test set and using the remainder for training and validation, resulting in 26,102, 4,105, and 10,996 test issues for the Apache, BTLink, and EasyLink datasets respectively. False links were generated using the RQ1 time frame (one year after issue creation and thirty days before and after issue closure). Since true issue-commit links are rare relative to the total number of commits in a repository, the resulting training data is heavily imbalanced. To address this, we applied under-sampling by selecting the top 10 commits retrieved by RRF for each training issue as negative examples. This strategy serves two purposes: it balances the class distribution by limiting the number of false links per issue, and it ensures that the negative examples are hard negatives, meaning commits that are semantically or temporally similar to the issue but are not true links. Training on such examples more closely mirrors the reranking task at inference time, where the model must distinguish true links from a set of plausible but incorrect candidates returned by the retriever.

We evaluated reranking techniques using $K=20$, meaning reranking techniques were given 20 commits per issue collected by retrieval methods. Since most issues contain only one or two true links, this means the reranker must identify and promote the correct commits from a candidate set containing approximately 18-19 false links, creating a challenging but realistic reranking scenario. A higher $K$ value would increase recall by capturing more true links in the candidate set, but would also introduce more noise and undermine the purpose of the initial retrieval stage, which exists precisely to reduce the commit search space to a manageable subset. Conversely, a very small $K$ risks excluding true links entirely, producing overly optimistic results that do not reflect real-world conditions. Thus, $K=20$ represents a practical balance: it is large enough to retain nearly all true links given our retrieval recall, while small enough to present a meaningful reranking challenge. 

We considered a number of techniques for the purpose of this evaluation. Feature-based approaches explicitly model signals that are known to correlate with issue-commit links, such as temporal proximity, developer overlap, and lexical similarity. FRLink uses TF-IDF vectors to score similarity between issue text and commit text providing a solid baseline~\cite{FRLink}. RCLinker relies on textual similarity and metadata features between issues and commits learned by a random forest classifier~\cite{RCLinker}. Hybrid-Linker combines a TF-IDF–based textual model with Gradient Boosting and a non-textual ensemble using Gradient Boosting and XGBoost to produce the final prediction~\cite{HybridLinker}. Deep learning-based approaches utilize embeddings to learn semantic patterns between issues and commits. Cross-encoders like ms-marco-MiniLM-L-6-v2 embed documents together, unlike bi-encoders that embed them separately, this captures fine-grained token-level interactions. BTLink combines two bi-encoders, RoBERTa for natural language and CodeBERT for code terms~\cite{BTLink}. For each approach, issue identifying information like the issue key is removed from the commit texts.

\begin{table*}[t]
\centering
\small
\setlength{\tabcolsep}{4pt}
\begin{tabular}{lccc|ccccc|cccccc}
\toprule
Reranker & Dataset & P@1 & R@1 & P@10 & Hit@10 & R@10 & MRR@10 & NDCG@10 &
P@20 & Hit@20 & R@20 & MRR@20 & NDCG@20 \\
\midrule
GPT5.1 & EasyLink & 0.822 & 0.709 & 0.118 & 0.911 & \textbf{0.878} & 0.863 & 0.852 & 0.060 & 0.916 & 0.884 & 0.863 & 0.853 \\
RCLinker & EasyLink & \textbf{0.842} & \textbf{0.733} & \textbf{0.119} & \textbf{0.912} & \textbf{0.878} & \textbf{0.871} & \textbf{0.858} & 0.060 & 0.916 & 0.884 & \textbf{0.871} & \textbf{0.858} \\
Hybird-Linker & EasyLink & 0.087 & 0.067 & 0.070 & 0.567 & 0.527 & 0.190 & 0.264 & 0.060 & 0.916 & 0.884 & 0.215 & 0.361 \\
FRLink & EasyLink & 0.598 & 0.522 & 0.112 & 0.879 & 0.843 & 0.696 & 0.721 & 0.060 & 0.916 & 0.884 & 0.699 & 0.733 \\
BTLink & EasyLink & 0.670 & 0.585 & 0.110 & 0.860 & 0.823 & 0.726 & 0.736 & 0.060 & 0.916 & 0.884 & 0.730 & 0.753 \\
Cross-Encoder & EasyLink & 0.807 & 0.704 & 0.115 & 0.906 & 0.867 & 0.842 & 0.831 & 0.060 & 0.916 & 0.884 & 0.843 & 0.835 \\
Llama & EasyLink & 0.639 & 0.550 & 0.107 & 0.834 & 0.793 & 0.710 & 0.716 & 0.060 & 0.916 & 0.884 & 0.716 & 0.739 \\
Qwen3 & EasyLink & 0.782 & 0.680 & 0.117 & 0.908 & 0.872 & 0.835 & 0.831 & 0.060 & 0.916 & 0.884 & 0.836 & 0.833 \\
Gemma & EasyLink & 0.573 & 0.495 & 0.103 & 0.811 & 0.769 & 0.657 & 0.670 & 0.060 & 0.916 & 0.884 & 0.664 & 0.700 \\

GPT5.1 & BTLink & 0.920 & 0.799 & 0.125 & 0.963 & 0.924 & 0.938 & 0.917 & 0.064 & 0.968 & 0.933 & 0.939 & 0.918 \\
RCLinker & BTLink & \textbf{0.936} & \textbf{0.809} & \textbf{0.126} & \textbf{0.967} & \textbf{0.929} & \textbf{0.949} & \textbf{0.926} & 0.064 & 0.968 & 0.933 & \textbf{0.949} & \textbf{0.926} \\
Hybird-Linker & BTLink & 0.201 & 0.173 & 0.086 & 0.703 & 0.652 & 0.322 & 0.390 & 0.064 & 0.968 & 0.933 & 0.340 & 0.467 \\
FRLink & BTLink & 0.791 & 0.677 & 0.124 & 0.957 & 0.916 & 0.850 & 0.848 & 0.064 & 0.968 & 0.933 & 0.851 & 0.851 \\
BTLink & BTLink & 0.622 & 0.547 & 0.111 & 0.897 & 0.850 & 0.702 & 0.724 & 0.064 & 0.968 & 0.933 & 0.707 & 0.747 \\
Cross-Encoder & BTLink & 0.829 & 0.726 & 0.121 & 0.956 & 0.912 & 0.871 & 0.862 & 0.064 & 0.968 & 0.933 & 0.871 & 0.868 \\
Llama & BTLink & 0.762 & 0.663 & 0.114 & 0.908 & 0.858 & 0.813 & 0.804 & 0.064 & 0.968 & 0.933 & 0.817 & 0.824 \\
Qwen3 & BTLink & 0.909 & 0.791 & 0.123 & 0.963 & 0.920 & 0.929 & 0.906 & 0.064 & 0.968 & 0.933 & 0.929 & 0.909 \\
Gemma & BTLink & 0.635 & 0.544 & 0.107 & 0.847 & 0.796 & 0.709 & 0.710 & 0.064 & 0.968 & 0.933 & 0.717 & 0.746 \\

GPT5.1 & Apache & \textbf{0.953} & \textbf{0.882} & \textbf{0.111} & 0.971 & 0.943 & \textbf{0.961} & \textbf{0.942} & 0.056 & 0.972 & 0.946 & \textbf{0.961} & \textbf{0.942} \\
RCLinker & Apache & 0.949 & 0.878 & \textbf{0.111} & \textbf{0.972} & \textbf{0.945} & 0.959 & 0.941 & 0.056 & 0.972 & 0.946 & 0.959 & 0.941 \\
Hybird-Linker & Apache & 0.162 & 0.148 & 0.072 & 0.647 & 0.613 & 0.277 & 0.349 & 0.056 & 0.972 & 0.946 & 0.299 & 0.437 \\
FRLink & Apache & 0.773 & 0.714 & 0.108 & 0.951 & 0.921 & 0.833 & 0.840 & 0.056 & 0.972 & 0.946 & 0.835 & 0.847 \\
BTLink & Apache & 0.896 & 0.829 & 0.110 & 0.964 & 0.935 & 0.920 & 0.909 & 0.056 & 0.972 & 0.944 & 0.920 & 0.911 \\
Cross-Encoder & Apache & 0.895 & 0.830 & 0.110 & 0.966 & 0.937 & 0.920 & 0.909 & 0.056 & 0.972 & 0.946 & 0.921 & 0.912 \\
Llama & Apache & 0.710 & 0.655 & 0.099 & 0.883 & 0.849 & 0.766 & 0.770 & 0.056 & 0.972 & 0.946 & 0.772 & 0.796 \\
Qwen3 & Apache & 0.909 & 0.841 & 0.110 & 0.969 & 0.938 & 0.931 & 0.917 & 0.056 & 0.972 & 0.946 & 0.932 & 0.919 \\
Gemma & Apache & 0.652 & 0.593 & 0.098 & 0.838 & 0.806 & 0.716 & 0.723 & 0.058 & 0.975 & 0.952 & 0.725 & 0.761 \\
\bottomrule
\end{tabular}
\caption{Performance of Different Reranking Techniques on our Datasets}
\label{table:rq3_results}
\vspace{-0.6cm}
\end{table*}

We also explored the use of LLMs for issue-commit link recovery. Unlike supervised reranking models, LLM-based approaches do not rely on task-specific training data but instead leverage general-purpose reasoning capabilities learned during pre-training. The flexibility and large range of tasks that can be completed by these models is attractive, and they are reported to perform well in many scenarios. Running these models locally can require powerful hardware. Alternatively, subscription-based APIs can be costly and have usage limits, which may limit their practical applicability. For our LLM-based pipeline, we use the same prompt as Huang et al. \cite{EasyLink} containing the concatenated issue title and description, as well as the commit ids and messages. The prompt instructs the model to re-rank the commits according to their relevance to the issue, returning only a list of ordered commit ids. The prompt can be found in our replication package. For each issue, we selected the top 20 RRF retrieved commits for our LLM reranking. Our chosen LLMs include GPT-5.1, Gemma-7B~\cite{gemmateam2024gemmaopenmodelsbased, gemma7b_it}, Qwen3-32B\cite{qwen3-32b,qwen3technicalreport}, and Llama-3.1-8B-Instruct~\cite{llama3_1_8b_instruct}.

\textbf{Results:} Our evaluation shows that the strongest reranking performance comes from RCLinker, GPT-5.1, Qwen3-32B, and our cross-encoder model as shown in Table~\ref{table:rq3_results}. Despite differences in design and architecture, these methods consistently rank true issue-commit links in the top positions. RCLinker leverages repository metadata and workflow signals, while our small cross-encoder model (ms-marco-MiniLM-L-6-v2) achieves results comparable to Qwen3-32B, highlighting its semantic understanding. Among LLMs, GPT-5.1 and Qwen3-32B perform best, whereas smaller models show more variable results (Table~\ref{table:rq3_results}). RCLinker consistently achieves the strongest performance across datasets, achieving precision@1 of 84.2\%, 93.6\% and 94.9\% across our datasets, outperforming both deep-learning and LLM-based techniques. Despite relying on a comparatively simple random forest classifier, RCLinker learns metadata features such as temporal distance between issue creation and commits, developer overlap between issue reporters and commit authors, and file-level indicators, as well as TF-IDF similarities between different textual data. The strong performance of RCLinker suggests that issue-commit link recovery is not solely a semantic matching task, but one that benefits significantly from incorporating contextual information about repository patterns and developer workflow. Our findings also highlight that simpler baselines remain highly competitive and, in several cases, outperform more complex models (Table~\ref{table:rq3_results}). This suggests that increased model complexity does not necessarily translate into superior reranking performance.

\section{Key Takeaways}\label{sec:takeways}
In information retrieval systems, filtering out irrelevant information is important. Our results have shown that issue closure time, can be an important feature for identifying commits that are likely to be linked to a given issue. We have found that dense retrievers consistently outperform sparse methods in retrieving relevant commits. However, combining dense and sparse methods through fusion methods such as reciprocal rank fusion can further enhance retrieval effectiveness.
From our study on reranking techniques, we find that reranking effectiveness in issue-commit linking depends strongly on technique choice. Traditional approaches (e.g., RCLinker), which leverage textual and metadata features can perform competitively with more complex techniques, while some models, such as Hybrid-Linker, may degrade reranking quality. Cross-encoders offer stable performance, and large language models show mixed results: larger models, such as GPT-5.1, perform well, whereas smaller models, such as Llama, are less effective. This highlights that increased technique complexity alone does not guarantee improved performance.
Overall, our findings emphasize that before adopting complex or resource-intensive techniques (e.g., LLMs), simpler approaches should be tested first, as they can deliver robust performance that may exceed the benefit of their complex alternatives.
\textbf{Computational Cost:} Evaluating GPT-5.1 across all 41,203 test issues cost \$128.10 USD in total, or approximately \$0.004 USD per issue, using OpenAI's batch API at a 50\% discount over standard pricing. While this per-issue cost is modest, it scales linearly with dataset size and may become a practical constraint for very large repositories or continuous integration settings where links must be recovered in real time. In contrast, RCLinker and the  Cross-Encoder can be trained in anywhere from under an hour to a few hours depending on the size of the repository, and evaluated in minutes, requiring only modest hardware and no subscription fee.


\section{Related Work}\label{sec:related_work}
A number of techniques have been developed to link commits to issues. Existing techniques can be divided into the following three categories.

\textbf{Heuristic-based Approaches:} There are many heuristic-based approaches to traceability link recovery. To establish the link between commits and bug reports, we can scan commit logs for reference to bug reports. However, the problem is that this can give us only a sample of links instead of the complete population since the inclusion of references to bug reports in commit messages is not enforced in software development \cite{Bachmann}. Thus, we are missing a number of links. For example, Bachmann et al. developed an interactive tool, called Linkster, that can allow developers to browse through bug reports and commit history and manually annotate the missing links \cite{Bachmann}. Wu et al. developed a technique, called ReLink \cite{Wu}, leveraging textual similarity between bug reports and changed logs, bug owners and change committer mapping, and bug-comment timing. Nguyen et al. introduced MLink \cite{Nguyen}, which combined textual and source code features of modified files to establish the bug-to-fix links. Schermann et al. leveraged a set of heuristics to establish the missing link \cite{Schermann}. However, the problem is that heuristic-based approaches often miss links~\cite{fair_and_balanced}.

\textbf{Machine Learning-based Approaches:} Missing traceability links make software systems harder to maintain. As a result, machine learning techniques have been widely used to automatically establish links between issues and commits. For example, Sun et al. \cite{FRLink} proposed FRLink, a model that leverages TF-IDF similarity between issues, commit messages, and file changes, and learns a threshold to achieve 90\% recall. Le et al. \cite{RCLinker} developed RCLinker, which trains a random forest using nine textual features and eleven metadata features to identify issue-commit links. A major challenge in this domain is the scarcity of positive labelled data, as negative links are often derived from positive ones, resulting in limited training data. To address this, Zhu et al. \cite{Zhu} proposed TRACEFUN, an approach that identifies unlabelled artifacts semantically similar to labelled ones and uses them to train two deep learning models, TraceBERT and TraceNN. Similarly, Sun et al. \cite{Sun} introduced PULink, which extracts eight features from issues and commits to train a model using both labelled and unlabelled links. Mazrae et al. \cite{HybridLinker} developed Hybrid-Linker, which trains separate classifiers on textual and non-textual features and combines their outputs into a final model. Dong et al. \cite{Dong} showed that state-of-the-art traceability models that perform well on open-source datasets often struggle in industrial settings due to class imbalance and a lack of labelled data. To mitigate these issues, they proposed SPLINT, a semi-supervised approach that generates pseudo-labels for unlabelled data to balance the training set. Finally, Rath et al. \cite{Rath} investigated which features most strongly indicate issue–commit links, leveraging commit time, resolution time, and textual similarity for link detection. While machine learning-based approaches have advanced link recovery, they heavily rely on labelled data and may miss links when data is sparse or noisy.

\textbf{Deep Learning and LLM-based Approaches:} In recent years, deep learning techniques have become more relevant in this space. Ruan et al. used word embedding and a recurrent neural network to learn the semantic representation of textual descriptions and code from issues and commits to establish their missing links \cite{Ruan}. Xie et al. developed a technique that utilized a knowledge graph to represent code context \cite{Xie}. The embeddings from the code knowledge graph, as well as those generated from issue and commit text, are used to link issues and commits using RNNs and SVMs. Lin et al. developed a BERT-based framework that used a three-step training approach to establish traceability links between issues and commits \cite{Lin}. Zhang et al. proposed a technique for issue-commit linking, called EALink. The technique utilized knowledge distillation from a pre-trained model to generate a compact model and contrastive learning to capture issue-commit correlation \cite{Zhang}. Lan et al. proposed BTLink, which leveraged a BERT-based technique and pre-trained models to establish missing issue-commit links using issue-text, commit-text, and commit-code \cite{BTLink}. Zhu et al. addressed the problem of limited label data by leveraging pre-trained models and deep semi-supervised learning to learn the pseudo labels of unlabeled data \cite{JZhu}. The most relevant work to our study is EasyLink, which used a combination of selection, retrieval, and reranking mechanisms to establish issue-commit linking \cite{EasyLink}. For each issue, the technique considers one year time window after issue creation to select plausible commits. The technique leverages an information retrieval mechanism to retrieve related commits from an issue, then applies an LLM to rank them. While the study established the importance of considering a longer time window to generate plausible links, it did not explore choices for selection, retrieval, and reranking.

\section{Threats to Validity}\label{sec:ThreatsToValidity}
This section discusses threats to the validity of our research.

\textbf{Threats to External Validity.} Threats to external validity concern the generalizability of our findings. Our research questions are answered using projects collected from open-source software repositories, which may limit the applicability of our results to other types of software systems. However, we mitigate this threat by using a dataset derived from real-world, widely used projects that rely on Jira for issue tracking. Unlike prior studies that reuse established benchmarks, we construct a new dataset and apply strict selection criteria to ensure that the included projects are actively maintained and representative of non-trivial, production-quality systems. Consequently, while our findings may not generalize to all software projects, they are likely to carry forward to similar large-scale, actively developed open-source systems that use issue tracking systems.

\textbf{Threats to Internal Validity.} Threats to internal validity refer to potential biases or errors in our research methodology. We constructed the dataset by linking issues to commits. We applied heuristics to establish the link between issues and commits. While we cannot guarantee that our dataset does not contain invalid links, we would like to emphasize that we took great care to establish the issue-commit link. Our dataset is highly imbalanced (i.e., the number of false links is significantly greater than the number of true links). However, we carefully reviewed all commits and divided the dataset into training and test sets, ensuring no overlap between the two.

\section{Conclusion}\label{sec:conclusion}
In this study, we proposed a realistic time frame to model issue-commit linking, capturing the characteristics observed in real-world software development. Using this setup, we evaluated six retrieval methods and found that dense vector-based approaches consistently outperformed sparse methods. Beyond retrieval, we assessed nine reranking models ranging from classical machine learning techniques to state-of-the-art large language models. Our results indicate that, while large language models offer strong performance, simpler machine learning models can achieve comparable effectiveness, demonstrating that increased model complexity does not automatically translate into better results. Overall, our findings suggest that testing baseline models is important before adopting high-complexity solutions. Future work should examine whether task-specific prompts can improve smaller open-source LLMs. It should also benchmark newer models, such as Claude and Gemini, against proprietary and traditional methods (e.g., RCLinker). Finally, improved negative sampling and learned fusion strategies could further enhance retrieval performance. The replication package and dataset for this study are publicly available at \cite{data}.

\section{Acknowledgements}
The authors gratefully acknowledge support from the Natural Sciences and Engineering Research Council of Canada (NSERC) through a Discovery Grant and the Undergraduate Student Research Awards (USRA) program. This research was enabled in part by support provided by the Digital Research Alliance of Canada (alliancecan.ca).


\bibliographystyle{ACM-Reference-Format}
\bibliography{references.bib}

\end{document}